# Intrusion Detection – A Text Mining Based Approach

**Gunupudi RajeshKumar[1], N Mangathayaru[2], G Narsimha[3]**

[1,2]Faculty of Information Technology, VNR VJIET, India
[3]Faculty of Computer Science and Engineering, JNT University, Jagityal, India

**Abstract:** Intrusion Detection is one of major threats for organization. The approach of intrusion detection using text processing has been one of research interests which is gaining significant importance from researchers. In text mining based approach for intrusion detection, system calls serve as source for mining and predicting possibility of intrusion or attack. When an application runs, there might be several system calls which are initiated in the background. These system calls form the strong basis and the deciding factor for intrusion detection. In this paper, we mainly discuss the approach for intrusion detection by designing a distance measure which is designed by taking into consideration the conventional Gaussian function and modified to suit the need for similarity function. A Framework for intrusion detection is also discussed as part of this research.

## 1. Introduction

Intrusion may be defined as an activity that essentially attempts to compromise integrity, authenticity, confidentiality, availability of system resources. For an organization to be safe, efficient and most reliable it must maintain several layers of security. These include network security and information security. This challenge is becoming much more complex currently as systems and services are becoming complex facilitating several new possibilities for attackers.

One may achieve information security by maintaining confidentiality, integrity and availability. Also as the data is enormously increasing and turning into big data, various design challenges, data analysis challenges, requirement for new algorithms, methodologies and measures are coining out. This further makes the situation more complex to handle.

Another problem which coins is the curse of dimensionality. Intrusion detection is not free from the problem of dimensionality and must be handled without fail for accurate results. The process called knowledge discovery from databases may be used in hand with the methods and methodologies of intrusion detection. Data Mining and Intrusion Detection go hand in hand now days.

Intrusion detection systems combine data mining methods, methodologies and algorithms along with the attack detection in to the system so that , the system can detect the intrusion dynamically. Similarly, the Intrusion detection systems (IDS) which mainly use anomaly detection mechanisms try to discover the abnormal behaviours.

In spite of several detection mechanisms available, there is a dearth of proper mechanism which can fix the behaviour as the intrusion. Though the findings and methods work for standard datasets but they ultimately fail over the real time datasets generated dynamically.

Figure 1 shows, the framework of NIDES intrusion detection system. We can also combine both the signature based and anomaly based intrusion techniques to forma hydrid technique to come up with the decision on normal and intrusive traffics. The figure 2 shows the knowledge discovery framework depicting each stage of the knowledge discovery process which may be used along with intrusion detection mechanisms to improve the efficiency and optimize the output results.

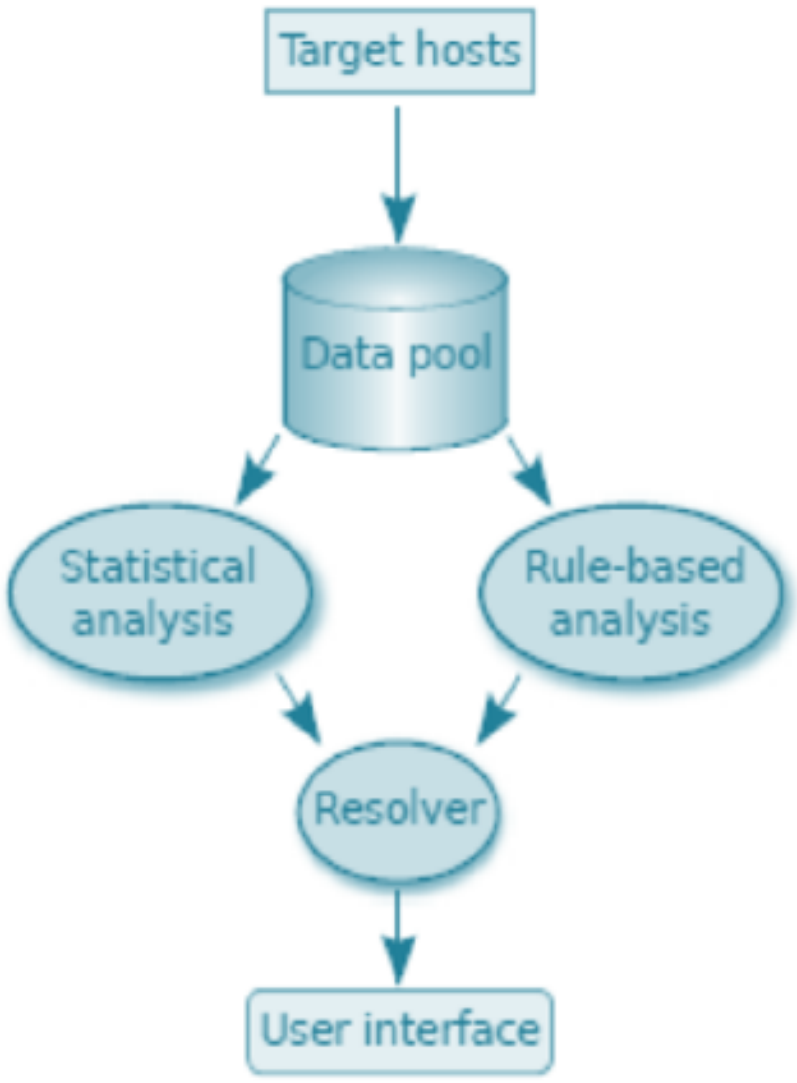

**Fig 1: Framework of NIDES Intrusion Detection**

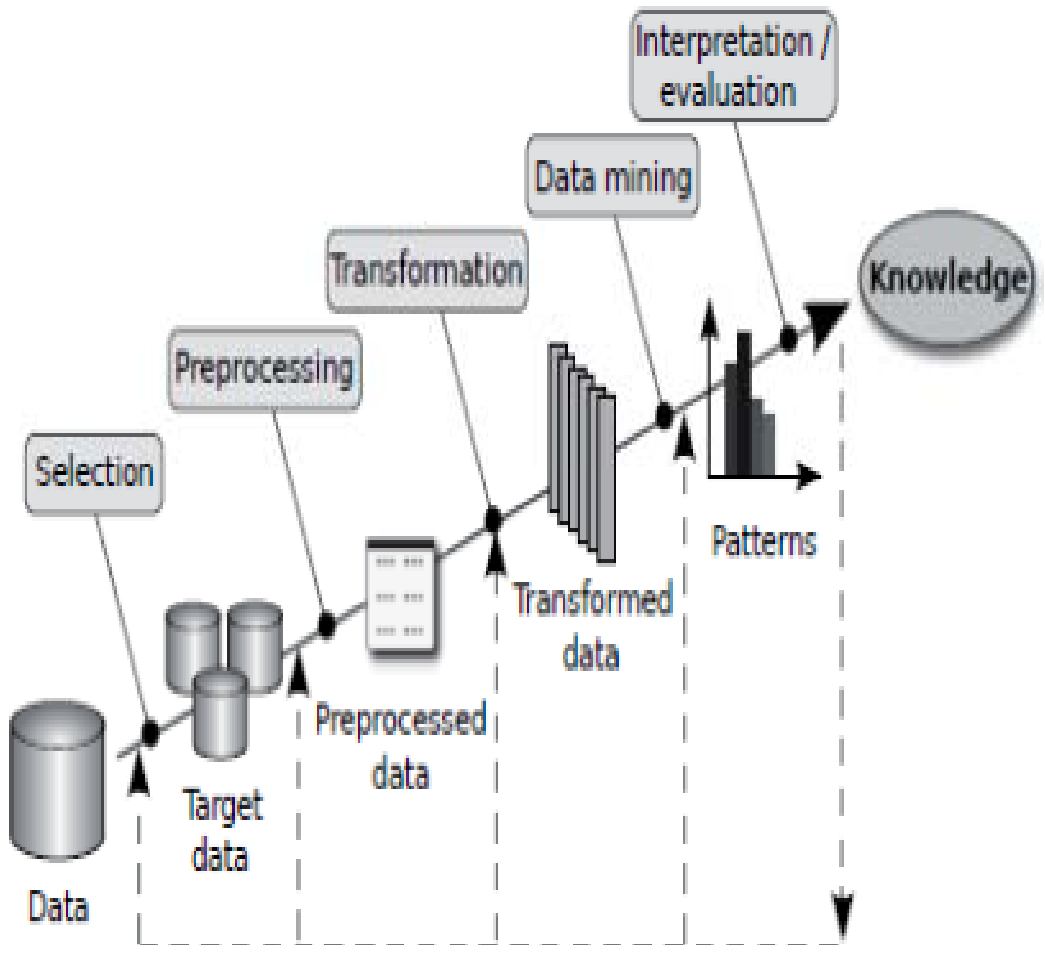

**Fig 2 : Framework of KDD process**

Securing internal and external resources for any organization in an unauthorized way, is becoming a major alarming problem now-a-days. Any sensitive information usually attracts attention from intruders leading resources to become vulnerable. The design of intrusion detection systems mainly functions based on two concepts. The former is the misuse based intrusion detection system and the later is the anomaly based intrusion detection system [26]. For the misuse based intrusion detection system, it is mandatory to construct a knowledge base which is useful in taking decisions whether the incoming request is normal or intrusion. The knowledge base is a collection of known signatures and instances of intrusion attacks.

Whenever a new request arrives, its signature will be cross checked with the already existing signatures in knowledge base and the decision is taken. Alert will be generated by the intrusion detection system, in case if it is a threat.

The second methodology is anomaly based, where the intrusion detection system learns the behavior of the system, and will immediately generate an alert in the case of deviation from the normal behavior [1,2,5].

Research contribution towards building IDS using different data mining techniques has been extensively studied in literature. In order to detect threats, it is advantageous to use soft computing techniques rather than traditional approaches for construction of the IDS.

## 2. Related Work

Intrusion detection system monitors all the incoming traffic and restricts entry of an unauthorized attempt to protect the resources by applying suitable rules. Recent research publications in intrusion detection algorithms concentrated more on the feature extraction from the data. The following is the list of various related techniques published in various journals for intrusion detection as shown in Fig. 1.

### 2.1 Text Processing

In this approach for intrusion detection, the system calls serve as the source for mining and predicting any chance of intrusion. When an application runs, there might be several system calls which are initiated in the background. These system calls form the basis and the deciding factor for intrusion detection [1, 2, and 3]. The approach of intrusion detection using text processing has been one of the research interests among researchers working in the area of network and information security.

This technique uses system call sequences [1] by applying text processing techniques. Alok Sharma et. al 2007, in his paper discussed the intrusion detection mechanism using text processing technique k-nearest neighbor (kNN) classifier. This classifier was used on the DARPA 1998 database and the results were proved to be better than other algorithms[2]. In their paper, Alok Sharma et. al 2007, demonstrated the cosine similarity measure and binary similarity measure.

For the first time Liao et. al[2] used the cosine similarity measure and later Rawat et. al. [3] extended by introducing the weight component for the system calls.

Alok Sharma et. al. proposed a new similarity measure which not only considers the frequency of the system calls rather than the number of common system calls between the processes.

**2.2 SVM**

SVM is one of the finest supervised methods used for classification. It includes learning algorithms through which, the training data gets classified. In order to get more quality in the classification process, SVM uses high dimension values for classification [12].

Initially, the training values are given through which these values are classified. SVM is generally used for the classification and Regression. In the process of determining the intrusion using system calls, if the number of system calls is too many, it will be difficult to perform the classification keeping performance unchanged. In order to have the performance of classification process unchanged, the dimensions need to be reduced without affecting the quality

**2.3 Signature Detection**

Intrusions are generally detected by matching captured pattern with already preconfigured knowledge base. The rate of false alarms in the case of unknown attacks is very high [25]. In his paper ,Yuxin Meng at.al. narrates that the Intrusion Detection System based on Signature [14] smells an attack by analyzing its stored signatures with information within the packet payloads. The signature may be a collection of rules which can be formed as an identity, which shall be stored in the database. In conclusion, the signature based intrusion detection is one of the prominent approaches to detect the threats, even though it has a drawback of possibility of occurrence of false alarms. The detection accuracy is high in the case of previously known attacks and computational cost is very less.

**2.4 Genetic Algorithms**

These GA techniques are generally used to select the best features that are used for IDS and when compared to other methods, has better efficiency. This is an approach which is, slightly trickery, complex and hence need to be used in specific manner rather in general approach [9, 23]. Studies based on both Genetic Algorithms and Fuzzy rule based systems can be classified as Michigan, Pittsburgh approaches. The Pittsburgh method uses a set of if-then fuzzy rules are coded as an individual. In Michigan only a single if-then fuzzy rule is coded as individual.

**2.5 Fuzzy Logic Approach**

The fuzzy logic approach considers the approximate theory rather than taking exact inference from predicate logic into consideration. These methods use quantitative features. This technique provides improved flexibility to some uncertain problems. However, when compared to the artificial neural network approach, the detection accuracy is lesser [25]. These fuzzy approaches can also be used for the anomaly detection as the features can also be considered as fuzzy variables [22]. As long as the observation lies within prescribed intervals, this kind of processing schemes are to be considered as normal (Dickerson, 2000)[23]. Sometimes intrusion detection systems based on the anomaly flags observed, activities that keep deviating from normal attribute patterns [20].

### 2.6 Anomaly Based Approach

These approaches use statistical test on the collected behavior to identify the occurrence of intrusion. Time taken to identify is more in this method. The detection accuracy is directly proportional to the collected amount of behavioral pattern features. The rate of false alarms is less in the case of unknown attacks [25]. Sang Hyun Oh, et. al. (2003) proposed in his paper [4]that an anomaly detection measure that uses a clustering algorithm which models the normal behaviors of users activities.

The statistical analysis predictive pattern generation and data mining techniques classify an anomaly detection model [5]. The values related to features of user activity represent the corresponding feature rate for the execution of the activity. Because of this, the value domain of the features is generally mentioned in the form of integers, makes association and sequence data mining inapplicable.

Hence only frequent item sets be found among the finite number of categorical items. In the contrary, this kind of problems can be better handled by the clustering as it is purely based on the similarity of data.

For analyzing, the common properties of all transactions of a user, anomaly based intrusion detection techniques are included in a host based IDS. The DBSCAN, JAM, ADMIT and EMERALD are few algorithms combined for Intrusion Detection Systems for ensuring successful threat detection. The anomaly based intrusion detection algorithms can be classified into three categories [21] as follows:

i. Statistical based : Nature of the process involved, is univariate, multivariate, time series model.

ii. Knowledge based: Nature of the process involved are FSM, Decryption languages, Expert systems.

iii. Machine Learning based: All soft computing techniques comes into this group

### 2.7 Association Rule Based

These techniques are used for only known attach signatures and/or relevant attacks in the misuse detection. Total unknown attacks are not at all detected and moreover it requires more number of database scans, to generate the rule base [25]. Lee, at.al. initiated the concept of using association rules for intruder detection solutions and was extended in [15, 16, 17].

### 2.8 Dimensionality Reduction

In order to avoid false alarms, two techniques need to be completed without fail. They are Preprocessing techniques and Dimensionality Reduction techniques. All the system calls that were captured by the Intrusion Detection System, many not play a role in deciding whether the incoming request is a threat or not. In this case, the system calls which were not important or irrelevant to the threat detection process must be identified and removed from the database. That is how a preprocessing system will decide each unit of data, whether it is a normal data or an anomalous data.

The preprocessing tasks involve activities like dataset creation, cleaning, integration, feature building. These most important steps to be discussed are:

***1. Dataset Creation:*** Data to be identified and collected in order to proceed for the preprocessing process. The data need to be

separately identified for Training Phase and Testing Phase.

***2. Feature Building:*** To improve the discriminative properties for the anomaly detection process, additional features are described for the data. This feature building can be done manually or using some automated tools.

## 3. Knowledge Discovery Approach

Most of the significant works carried for finding intrusion detection may be classified into the following classes

1. Machine Learning Based Approach
2. Unsupervised Learning Based Approach
3. Supervised Learning Based Approach
4. Genetic programming Based Approach

### 3.1 Machine Learning Approach

Machine learning is a self learning approach which requires a formal system which can update itself continuously each time the new data is generated and added to the system. In essence, it must be an autonomous system which can address the continuous changes coined out and integrate the knowledge database.

This process requires ability to learn from experience, analytical capability, self learning capability, ability to handle dynamic changes so as to be self updated.

In essence, the major task in the machine learning algorithms is to design, analyze, develop, and implement various algorithms and methodologies which guide the machines (computer systems) to gain the self learning capability. Machine learning may be classified into supervised and unsupervised learning techniques [20].

### 3.2 Supervised Learning Approach

In this approach for intrusion detection, we must know the class label to build the knowledge database or knowledge rules. This is because of this reason; we call it as supervised learning technique or classification.

Given a dataset, we split the dataset into training and testing sets and build the knowledge using the training set. Then we use the samples from the testing set to test the class label of the test case chosen from the testing test. In short, the task of supervised learning is to build a classifier which can effectively approximate the mapping between input and output samples of training.

Once we build a classifier, this is followed by measuring the classification accuracy. Classification requires choosing an appropriate function which can estimate the class label. This is followed by measuring the classification accuracy.

The most popular classifiers include Decision tree based Classifier; ANN based classifier, KNN Classifier, SVM Classifier. The simplest non-parameter classifier is the KNN-classifier which is used to estimate the class label of the test input by assigning the label of the nearest neighbor.

### 3.3 Unsupervised Learning Technique

In the intrusion detection based on supervised learning technique, we do not have any knowledge on the class labels of the input dataset. In such a situation, we aim to choose the classifier based unsupervised learning. This process is also called as clustering process. In unsupervised learning based technique the objective is to obtain a disjoint set of groups consisting of similar input objects. These groups may be used to perform decision making, to predict the future inputs.

The K-means clustering method is the most popular among the various clustering algorithms where k indicates the number of clusters to be formed from the input dataset. The K-means algorithm requires specifying the number of clusters to be formed well ahead.

In [20] the authors make use of this property to decide the number of clusters in their approach for intrusion detection.

### 3.4 Genetic Programming Approach

GA techniques are generally used to select best features that are used for IDS and when compared to other methods have a better efficiency. This approach is slightly tricky and complex and hence need to be used in specific manner rather in general approach.

Studies based on both Genetic Algorithms and Fuzzy rule based systems can be classified as Michigan, Pittsburgh approaches. The Pittsburgh method uses a set of if-then fuzzy rules which are coded as individual. In Michigan only a single if-then fuzzy rule is coded as individual.

## 4. Research Issues

The computation problems may be classified into two types. These include 1. Optimization Problems and 2. Decision problems. In optimization based problems, the objective is to aim for efficiency. In the decision based problems, we must output the decision as yes or no, true or false, etc. Intrusion detection may be considered as the decision problem where we need to classify if the target is an intrusion or not.

One approach of intrusion detection which is recently being concentrated is using text mining techniques. Data mining is a knowledge discovery process aiming at retrieving the unknown hidden information available, but not yet been identified and focused to derive important conclusions and findings. Intrusion detection and data mining have been complementing each other in research works performed by various researchers towards finding various possibilities, approaches to detect intrusion.

The data mining approaches such as prediction, classification, clustering, noise elimination have been extensively used in the intrusion detection process as discussed in the related works of section 2. In this section, our objective is to outline a generalized method for intrusion detection. The problem of predicting intrusion detection has been a major challenge for researchers from the medical domain as well as from the other fields of engineering such as health informatics, medical informatics and information retrieval. We now try to point out the various research issues in handling data sets.

### 4.1 Pre-processing Datasets

The research should first start with the study of the benchmark datasets. Sometimes there may be a need to start collecting data from scratch if we are working over a problem in a particular domain. Preprocessing phase is an essential phase to make the dataset suitable for process effectively to obtain accurate, efficient results by applying the newly designed method or already existing algorithm.

Since there is no specific standard dataset for intrusion detection, we choose to consider the KDD-Cup 99 dataset as the one considered in [20]. This contains 494,020 samples totally. The dimensionality of each data sample is 41. Of these 41 dimensions, a total of 9 are intrinsic type, 13 are content type while all the remaining 19 are traffic type. Each data sample of the dataset is classified in to 5 classes.

There are four types of attacks and one normal traffic class. Since the number of classes is five, this is basically as 5-Class Classification problem. Similarly we may choose to use the DARPA 1998 or 1999 standard dataset. In particular, the research should first start with the studying the benchmark datasets.

Sometimes there may be a need to start collecting data from scratch if we are working over a problem in particular domain. Preprocessing phase is an essential phase to make the dataset suitable for processing and handling effectively to obtain accurate, efficient results by applying the newly designed method or already existing algorithm.

### 4.2 Dimensionality Reduction

This phase includes extracting features from the dataset. These include feature selection, feature extraction, information gain, the application of linear discriminant analysis, noise elimination, dimensionality reduction by computing frequent patterns, etc. some of the recent works include application of text processing approaches for intrusion detection.

### 4.3 Distance Measure

The choice of distance measure is an essential task in prediction and classification processes. Some distance measures use the notation of vectors and other distance measures use non-vectors as input.

Some of the well-known distance measures include cosine distance measure, Manhattan distance, Euclidean distance, Jaccard measure. If the input is a frequency vector, we may use cosine measure for finding distance between the same. Alternately we may design our own measure to compute the distance between any two input entities.

### 4.4 Classification and Prediction Algorithm

The underlying dataset is the deciding factor for the choice of the algorithm. A single classification algorithm is not suitable for every dataset. Choosing an efficient classification method followed by inefficient distance measure may lead to improper estimation of intrusion prediction.

The existing classification algorithms have their own advantages and disadvantages, which need to be studied and chosen effectively.

### 4.5 Noise Elimination

In text mining based intrusion detection, we may have to form the process vs system call matrix for intrusion detection after finding the system call vector which contains all system calls. Since the dimensionality of the system call vector makes the dimensionality of system call matrix large, we may have to reduce the dimensionality.

After deciding the number of system calls, there may be one or more system calls which may be not important and may be discarded without any loss of information. Every effort must be made in this direction, so that the system call attributes which are of the least importance and insignificant affect may be eliminated.

## 5. Proposed Approach

The approach of intrusion detection using the text processing has been one of the research interests among researchers working in the area of network and information security. The proposed approach for intrusion detection is based on the concept of text processing and use of data mining techniques in the prediction and classification of intrusion.

Our intrusion detection is based on system calls. Formally, we treat the algorithm to be a function of system calls. In this approach for intrusion detection, the system calls serve as an important source for mining and predicting any chance of intrusion. When an application runs, there might be several system calls which are initiated in the background these system calls form the basis and the deciding factor for intrusion detection

### 5.1 Steps involved

The block schematic of the proposed approach is given in the figure 2 below. The following are the sequence of steps

#### 5.1.1 Stage 1

The DARPA Dataset is used, as it is publicly available, labeled and pre-processed ready for use. Preprocessing of the dataset is to make it suitable for use by the data mining algorithm and techniques used to handle the data.

#### 5.1.2 Stage 2

Perform dimensionality reduction of system calls, as all the system calls need not be important. We must identify those system calls which are not dominant and eliminate such system calls. The Singular Value Decomposition (SVD) technique may be used to perform the dimensionality reduction at this stage. By applying SVD, we can figure out most dominant and least dominant system calls. Such system calls which do not make any significant effect may be eliminated. All such system calls are called outliers. A simple thumb rule is to eliminate all the systems calls whose Eigen values are less than 1.

**5.1.3 Stage 3**

This stage involves deciding which system calls must be retained from the most dominant system calls obtained in the previous stage. A simple thumb rule is to consider all those system calls, which add up to 90% energy.

**5.1.4 Stage 4**

We may apply frequent pattern approaches for finding frequent system calls. In this case, we are trying to find the system call item sets. Finding system calls sets may be used to derive important association rules which may be helpful in performing classification and for predicting the trends among system calls. There is a scope for research in this direction as very less work is carried out.

**5.1.5 Stage 5**

This stage involves using suitable distance measure such Euclidean, Cosine, etc, Alternately, one may design his/her own kernel measure which may be used to perform classification. Such a distance measure which is designed must satisfy all the basic properties of the distance function [27].

**5.1.6 Stage 5**

The next stage involves the process of classification. There are less number of options available for the datasets. This process becomes much simpler as the DARPA dataset is used for this purpose. We may use DARPA dataset, as it is publicly available, preprocessed and ready for use. To perform this process there are two approaches, kernel based and distance based measures. Similarity measures such as the Cosine measure and the Jaccard measure such as various distance measures may be used. In the case of binary matrix representation of the system calls, the Jaccard distance measure may be used. The Cosine measure is used for the frequency based system calls. On the other hand, we may also use methods such as SVM classification.

Alternatively, the user may design a new kernel measure and use with SVM classifier to perform classification.

**5.1.7 Research Direction**

There is scope for research, if we make use of association rules to perform dimensionality reduction. Efforts are countable in this direction as very less work is performed by researchers. One way is to find the relation between the system calls and reduce the system calls which are not important, if we already know the class as intrusion and non-intrusion.

**6. Text Mining Based Intrusion Detection**

The consensus based computing approach has been applied in various application areas which aims at using more than one algorithm or procedure, distance measures to address the respective problems. Since the chosen dataset has already defined the number of classes, and the intrusion detection is also a classification problem, we may choose to cluster the chosen dataset into number of clusters equal to the number of class labels.

In this paper, the objective is to use the K-means clustering method to cluster the chosen dataset into a number of clusters equal to the number of class labels. We may directly cluster the training set or alternatively choose perform feature selection followed by dimensionality reduction and then apply K-means clustering over this reduced dimensionality.

**6.1 Handling Training Set**

We follow the approach in [20] for dimensionality reduction. However, instead of using the conventional k-means algorithm, we choose to apply the modified K-means algorithm which uses the Gaussian based distance measure to find the similarity between data samples when forming the clusters. This is where the novelty of our approach starts with. In this approach, we reduce the dimensionality of training set by first applying a suitable clustering to a number of clusters equal to number of known class labels. Since the intrusions datasets are have labeled attacks, we can decide the number of clusters to be obtained. The better choice is k-means clustering algorithm as it clusters the input to the predefined number of clusters.

After, obtaining the clusters the next step is to find the distance between each training data sample and all the cluster centers. This is the first distance value computed. In addition to this for every data sample with in a cluster, we find its nearest neighbor within that cluster by selecting the pair of minimum distance. This is the second distance value.

The two distances are added to get a single distance. Now each data sample is mapped to a single distance value instead of data sample expressed as a function of system call attributes when performing text mining based intrusion detection. For example, if we consider the purpose of clustering, we must specify the number of clusters equal

**6.2 Distance Measure for K-means**

In this section, we discuss the distance measure used as part of the k-means clustering algorithm. We use the Gaussian function as the distance measure to find the distance between any two samples of training set. This may also be used to find the distance between any two data samples in general.

### 6.2.1 Gaussian Function

We consider the Gaussian function based distance measure to find the similarity between the data samples of the intrusion dataset. We use the same distance measure and apply the k-means algorithm to cluster the data samples.

For the purpose of dimensionality reduction, we use the k-means clustering technique to obtain the clusters using the proposed distance function and then find the distance between each training data sample and each of the cluster centroids. This is further followed by finding the nearest neighbor for every data sample within the cluster. These two distances are summed to get a new distance value. This distance value becomes singleton feature for each training data sample. Thus each data sample of the training set is mapped to a single feature value reducing the dimensionality to 1.

The Proposed distance function is defined as given by Equation. 1

$$G(x, \mu, \sigma) = \begin{cases} e^{-\left(\frac{x-\mu}{\sigma}\right)^2} & ; \quad \textit{one or both system calls exist} \\ 0 & ; \quad \textit{none of the system calls exist} \end{cases} \tag{1}$$

where
x = system call being considered
μ = mean of the system call w.r.t data samples present in the cluster
$\sigma$ = standard deviation of system call considered w.r.t data samples of training set.

The denominator of IDSIM is given by Equation.2 as shown below

$$H(x, \mu, \sigma) = \begin{cases} 1 & ; \quad \textit{one or both system calls exist} \\ 0 & ; \quad \textit{none of the system calls exist} \end{cases} \tag{2}$$

The average distance is the ratio of the two functions $G(x, \mu, \sigma)$ and $H(x, \mu, \sigma)$ and is formally represented as given by Eq.3

$$F_{avg} = \frac{G(x,\mu,\sigma)}{H(x,\mu,\sigma)} \tag{3}$$

The average distance considering distribution of all features hence is defined as Equation.4 as given below

$$F_{avg} = \frac{\sum_{i=1}^{i=n} 1 \sum_{s=1}^{s=m} e^{-(\frac{x_{is}-\mu_{is}}{\sigma_s})^2}}{\sum_{i=1}^{i=n} 1 \sum_{s=1}^{s=m} 1} \tag{4}$$

The distance function is represented as given by

$$IDSIM = (1+F_{avg}) / 2 \tag{5}$$

Where i indicates the $i^{th}$ data sample. S indicates the system call. IDSIM indicates the similarity function

### 6.3 Dimensionality Reduction of Training Set

Figure.3 shows the proposed approach for reducing the dimensionality of the training set and Figure.4 shows the proposed approach for reducing the dimensionality of the testing set using the proposed measure with K-means clustering technique.

So, we have both the testing and training sets with each data sample transformed to a singleton feature value. The test dataset can now be compared with training dataset in a very simple and effective, efficient way. The Proposed approach concentrates on using the Gaussian function based distance along with the K-means instead of conventional distance function used by K-means algorithm.

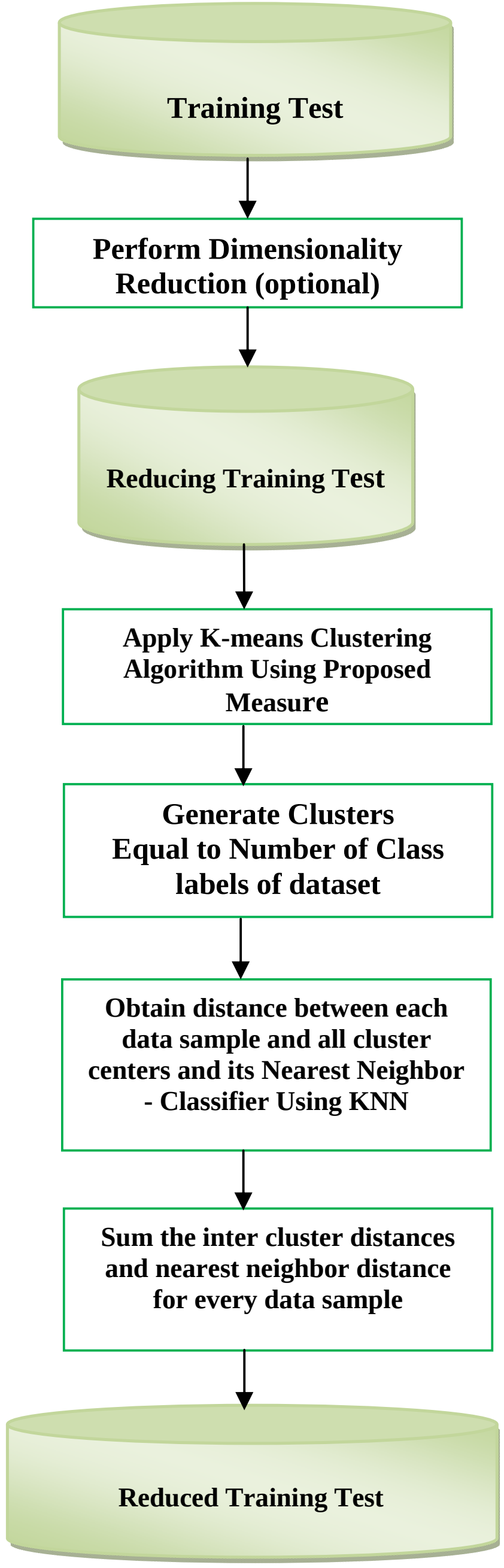

**Fig 3: Dimensionality Reduction of Training Set**

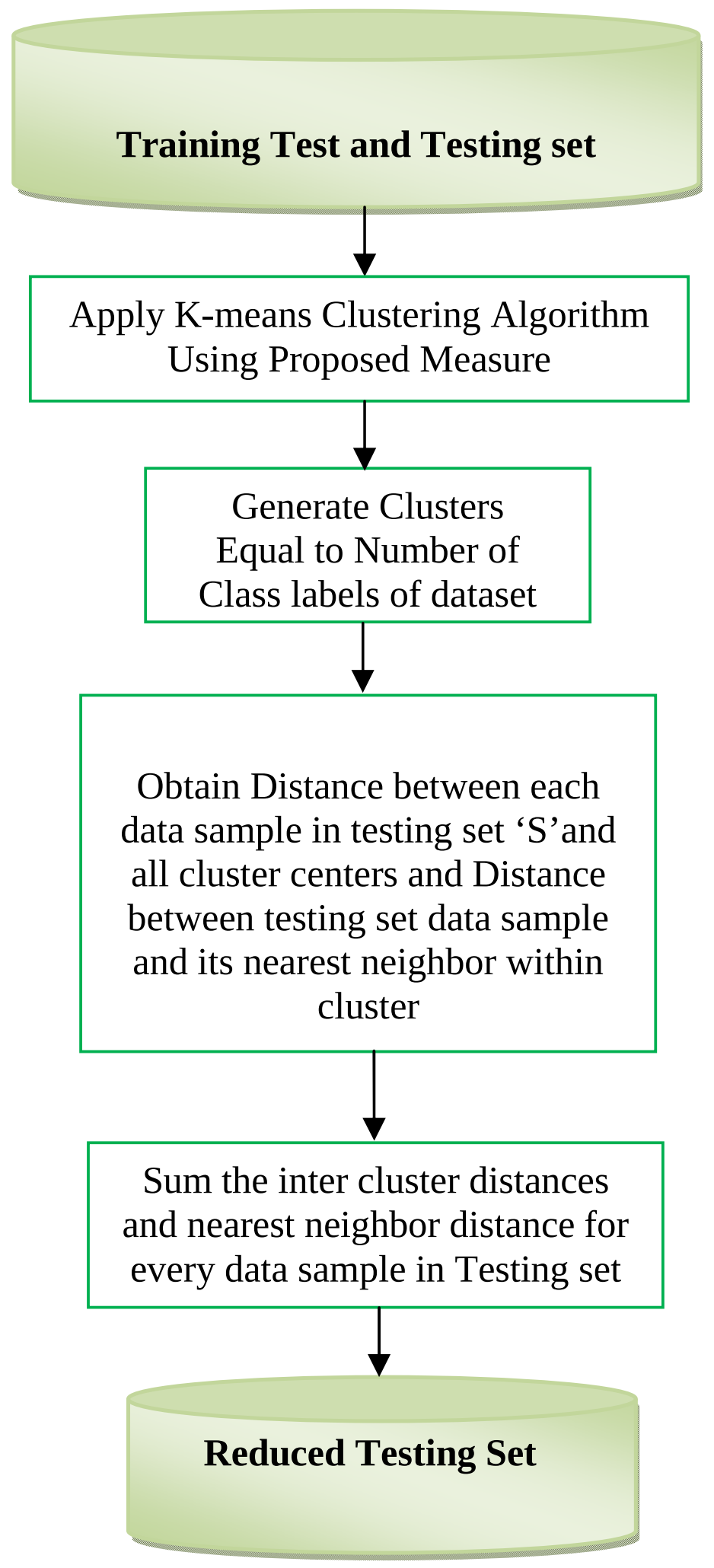

**Fig 4: Dimensionality Reduction of Testing set**

## 7. Conclusions

This paper discusses various approaches to be followed for detection of the intrusion. It also discusses the research issues to be considered in intrusion detection using text processing. It also discusses the sequence of steps to be followed in order to improve the effectiveness and performance of the intrusion detection mechanism and decreasing false alarms that are generated by the intrusion detection system. This paper also discusses the importance of preprocessing techniques, dimensionality reduction in order to reduce the false alarms. In this work, the second major contribution is in defining the similarity measure which has finite lower and upper bounds. The measure designed is Gaussian function based distance measure. The K-means algorithm is chosen for clustering using the proposed distance measure to cluster both the training and testing data samples. The training and test datasets are transformed to single dimensional feature with the use of k-means and proposed distance measure. The significance of the proposed distance measure is it considers the distribution of the system calls behavior over the entire training samples. This makes the computation accurate, even in binary form. The similarity value lies between 0 and 1.